\documentclass{article}
\usepackage{spconf,amsmath,graphicx}

\usepackage{subfig}
\usepackage[repeatunits=power]{siunitx}
\usepackage{pgfplots}
\usepackage{tikzscale}
\usepackage{upgreek}
\title{HIGH-SPEED MARKERLESS TISSUE MOTION TRACKING USING VOLUMETRIC OPTICAL COHERENCE TOMOGRAPHY IMAGES}
\name{M.~Schl\"{u}ter, L.~Glandorf, J.~Sprenger, M.~Gromniak, M.~Neidhardt, T.~Saathoff, and A.~Schlaefer}
\address{Institute of Medical Technology, Hamburg University of Technology, Hamburg, Germany}
\begin{document}
%
\maketitle
\begin{abstract}
Modern optical coherence tomography (OCT) devices provide volumetric images with micrometer-scale spatial resolution and a temporal resolution beyond video rate. In this work, we analyze an OCT-based prototypical tracking system which processes \num{831} volumes per second, estimates translational motion, and automatically adjusts the field-of-view, which has a size of few millimeters, to follow a sample even along larger distances. The adjustment is realized by two galvo mirrors and a motorized reference arm, such that no mechanical movement of the scanning setup is necessary. Without requiring a marker or any other knowledge about the sample, we demonstrate that reliable tracking of velocities up to \SI{25}{\milli\meter\per\second} is possible with mean tracking errors in the order of \SI{0.25}{\milli\meter}. Further, we report successful tracking of lateral velocities up to \SI{70}{\milli\meter\per\second} with errors below \SI{0.3}{\milli\meter}.
\end{abstract}
\section{Introduction}

Surgical interventions often require information about the pose of instruments relative to target regions. Examples include navigated procedures and active motion compensation \cite{van2015markerless, lin2019trajectory}. State-of-the-art tracking methods are commonly based on optical or electromagnetic systems \cite{maybody2013overview}. While marker-based tracking methods are widely used, they require a setup of the markers which may complicate the workflow \cite{busam2018markerless}. Especially in minimally invasive surgery, the placement of markers onto tissue structures is often not readily feasible. Hence, markerless approaches relying on natural features have been proposed. More conventional 3D imaging systems, like time-of-flight cameras or stereo vision, enable tracking of surface structures but often fail for smooth and homogeneous surfaces \cite{schaller2009time, ji2013tracking}. 

 Recently, the feasibility of a markerless tracking system based on optical coherence tomography (OCT) has been shown \cite{Schlueter19a}. OCT is a non-invasive imaging modality based on interferometric measurements and providing sub-surface information of scattering tissue. Usually, near-infrared light is used to acquire one-dimensional depth profiles of the tissue (A-scans) with a spatial resolution in the order of \SI{10}{\micro\meter}. By using scanning mechanisms, cross-sectional images (B-scans) and volumetric images (C-scans) can be obtained. Current systems operate with A-scan rates above \SI{1}{\mega\hertz} which allow for video-rate volume acquisition \cite{kolb2019live}. However, the size of the field of view (FOV) is typically only a few millimeters in each direction.
The additional sub-surface information delivers valuable input for tracking. In contrast to other imaging modalities used for intraoperative tracking, neither radiation exposure nor restrictions in the choice of surgical instruments due to magnetic fields limit the use of intrasurgical OCT \cite{kral2013comparison}. 

While OCT has been used for direct image guidance in ophthalmology, general motion tracking has not yet been widely studied. Laves et al.\ \cite{laves2017feature} investigated motion tracking on maximum intensity projections of 4D OCT images. A marker-based tracking approach using volumetric OCT was realized based on a 3D convolutional neural network to predict the 6D pose of a marker that can be attached to objects of interest \cite{gessert2018deep}. 

In this work, we extend a recently proposed setup for lateral motion tracking \cite{Schlueter19a}. First, we add a motorized reference arm to compensate for axial motion. Thereby, we are able to adjust the spatial FOV position to compensate for any translational motion of a sample. Second, we implement an efficient processing pipeline allowing for online processing at a volume rate of \SI{831}{\hertz}. This processing includes OCT image reconstruction and template matching to realize the tracking. Third, we present a setup to evaluate the 3D tracking performance of our system. Using both a 3D-printed phantom and tissue samples, we demonstrate that flexible and reliable markerless tracking of three-dimensional translational motion can be realized.

\section{MATERIAL AND METHODS}

\subsection{Hardware setup}
The basis of our system is a \SI{1315}{\nano\meter} swept-source OCT device (OMES, Optores, Germany) with \SI{1.59}{\mega\hertz} A-scan rate. We use a scan head based on a resonant galvo mirror, resulting in a C-scan rate of \SI{831}{\hertz} for small volumes of \num{32 x 32} A-scans. Hence, each image volume is of size \SI{32x32x480}{} along the lateral and depth directions, respectively. The FOV of our C-scans is approximately \SI{2.5x2.5x3.5}{\milli\meter}. Subsequently, the terms \textit{lateral} and \textit{axial} direction refer to the lateral and axial directions of the OCT image data. 

\begin{figure}[tb!]
    \centering
    
    \begin{tikzpicture}\small
    \node[inner sep=0] at (0,0) {\includegraphics[width=0.9\linewidth]{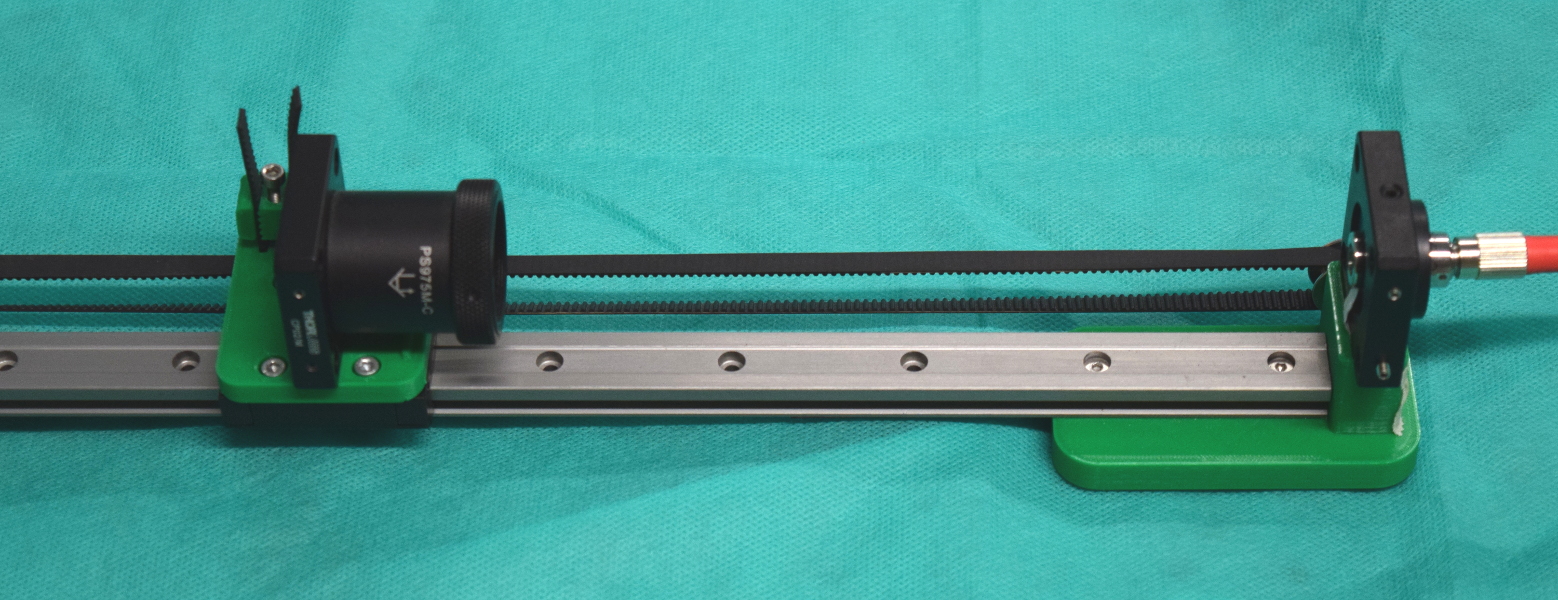}};
    \draw[|-|,black,thick,dashed] (-1.8,1) -- (2.9,1) node[midway,above,black,fill=white,circle,inner sep=0.6, yshift=2]{1};
    \node[fill=white,circle,inner sep=0.6] at (-2,-0.75) {2};
    \node[fill=white,circle,inner sep=0.6] at (3,-0.75) {3};
    \end{tikzpicture}
    \caption{The OCT's FOV is shifted axially by changing the distance (1) between the motorized reference mirror (2) and the fiber collimator (3).} 
    \label{f:refarm}
\end{figure}
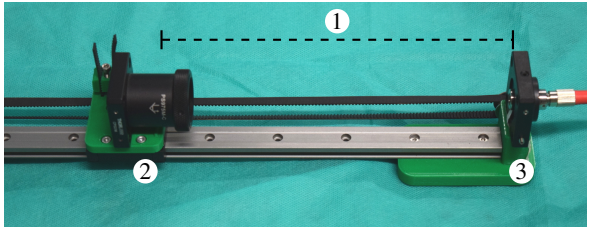

To follow a target which moves laterally, we use a second stage with two galvo mirrors \cite{Schlueter19a}. It allows to re-position the whole C-scan laterally without requiring any movement of the scan head. In order to adapt to axial motion of the target, we designed the motorized reference arm shown in Fig.~\ref{f:refarm}. By changing the distance between the fiber collimator and the reference mirror, the axial position of the C-scan is shifted. The maximum range is \SI{420}{\milli\meter} and a stepper motor is used to adjust the reference length with a step size of \SI{12.5}{\micro\meter} and a top speed of \SI{190}{\milli\meter\per\second}.

A simple achromatic lens with a focal distance of \SI{300}{\milli\meter} focuses the beam in the sample arm. This allows for a reasonable distance between the tracking system and the target. Furthermore, it provides acceptable image quality over a range of several centimeters. Note, that the actual image quality is not critical as long as structural patterns in the image volume can still be registered. The diameter of the lens is \SI{50.8}{\milli\meter} and practically limits the maximum lateral re-positioning range for the C-scan.

\subsection{OCT data processing}
We implemented a processing pipeline in CUDA which allows for online processing of all \num{831} volumes which we acquire per second. The pipeline covers both the image reconstruction and the image analysis. For this reason, we use a system with two GPUs. The raw OCT data are transferred to the first GPU (GeForce GTX TITAN X, Nvidia, USA) and are reconstructed as usual by subtraction of the background signal, re-sampling in $k$-space, windowing, Fourier transform with Nvidia's cuFFT library, and compression of the resulting intensity values. Subsequently, we use the phase correlation algorithm on the second GPU (GeForce GTX 980 Ti, Nvidia, USA) to determine the translation of each OCT volume with respect to an initial template volume. Phase correlation computes the translation which maximizes the cross correlation of two volumes. However, it uses a Fourier-domain approach and thereby mainly requires two fast Fourier transforms and element-wise operations on the data, which can be efficiently computed in parallel. It finally employs an inverse Fourier transform and the result has an intensity peak at the position corresponding to the translation. The search for this peak can be  parallelized by employing the reduction design pattern. We extended the basic algorithm by low-pass filtering of the cross-spectrum which reduces the noise in the results. This algorithm does not provide sub-pixel translations. Note that we use \num{10} volumes per second for online visualization while the remaining \num{821} volumes are used for tracking. 

From the phase correlation, we obtain the translations in image voxels. By converting them to steps of our galvos and our stepper motor and applying a proportional gain to avoid overshooting, we adjust the system parameters such that the OCT's FOV is centered at the template again. As very small oscillating movements would result in frequent change of direction of the stepper motor, we employ a threshold on the motor steps corresponding to a minimum traveling distance of \SI{75}{\micro\meter}.

\subsection{Experiments}
\begin{figure}[tb!]
    \centering
    \subfloat[\label{f:phantoms:plate}]{
    \begin{tikzpicture}\small
        \node[inner sep=0] at (0,0) {\includegraphics[trim=28 0 44 0,clip,width=.42\linewidth]{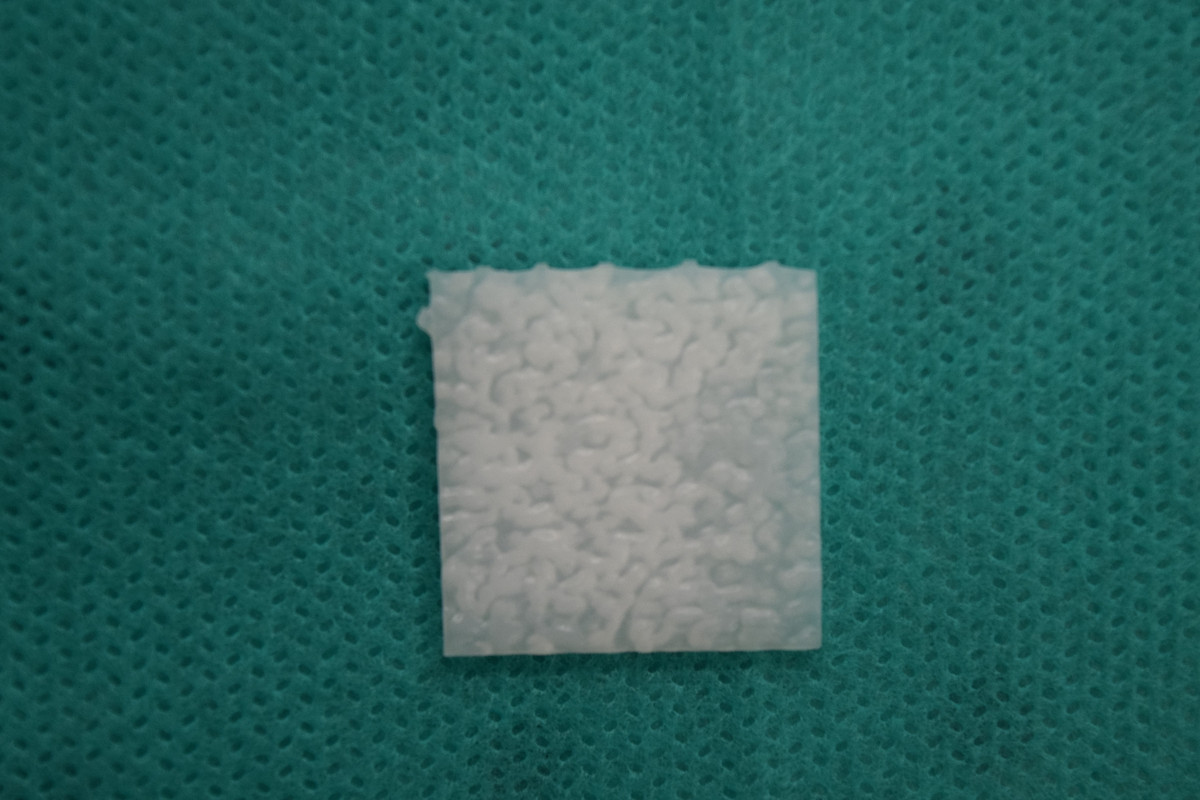}};
        \draw[|-|,white,thick] (-0.55,0.8) -- (1.05,0.8) node[midway,above]{\SI{18}{\milli\meter}};
        \draw[|-|,white,thick] (-0.85,0.55) -- (-0.85,-1.05) node[midway,xshift=-7,rotate=90,anchor=center]{\SI{18}{\milli\meter}};
    \end{tikzpicture}
    }\subfloat[\label{f:phantoms:turkey}]{
    \includegraphics[width=.42\linewidth]{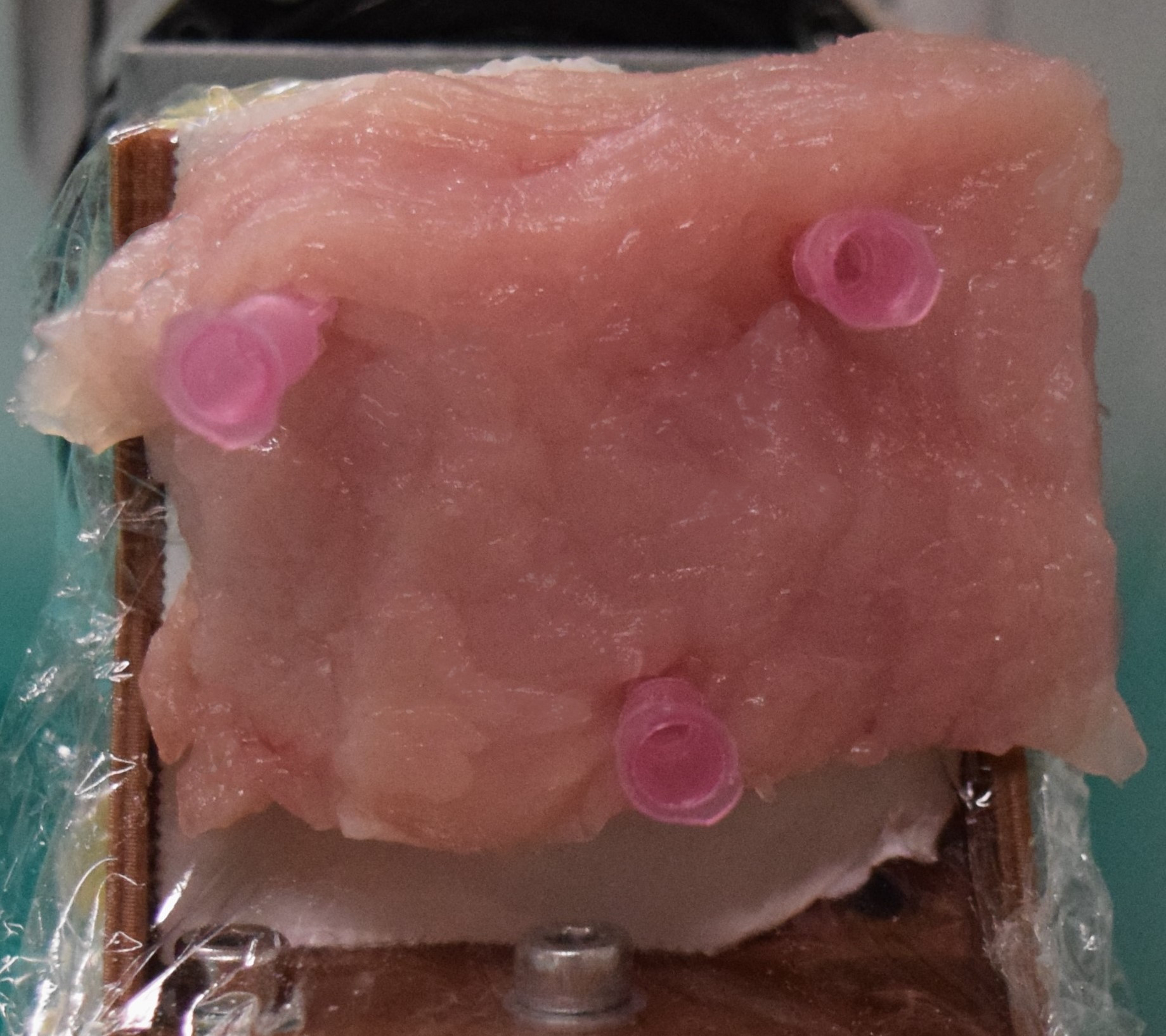}
    }
    \caption{Plate phantom (a) and exemplary turkey breast sample pinned to our sample holder (b)}
    \label{f:phantoms}
\end{figure}

To evaluate the range of velocities which we are able to track and the accuracy of our system, we use two different types of samples. The first one is a 3D-printed thin plate with a surface structure which shows random variations in the order of a few hundred micrometers (Fig.~\ref{f:phantoms:plate}). Additionally, we use samples of turkey breast (Fig.~\ref{f:phantoms:turkey}). We do not track specific areas of the samples, but simply position some part of them within the OCT's FOV. This part is then used as the template volume for subsequent tracking. 

\begin{figure}[tb!]
    \centering
    \begin{tikzpicture}\small
    \node[inner sep=0] at (0,0) {\includegraphics[width=.9\linewidth]{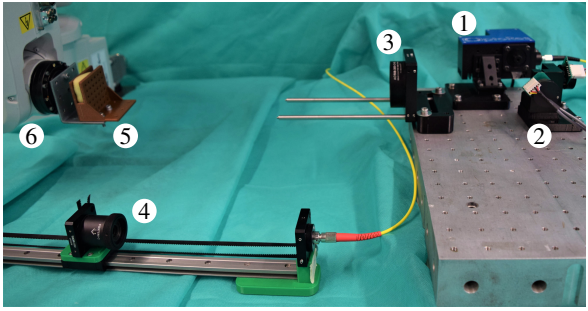}};
    \node[fill=white,circle,inner sep=0.6] at (2.25,1.75) {1};
    \node[fill=white,circle,inner sep=0.6] at (3.25,0.25) {2};
    \node[fill=white,circle,inner sep=0.6] at (1.25,1.5) {3};
    \node[fill=white,circle,inner sep=0.6] at (-2,-0.75) {4};
    \node[fill=white,circle,inner sep=0.6] at (-2.25,0.25) {5};
    \node[fill=white,circle,inner sep=0.6] at (-3.5,0.25) {6};
    \end{tikzpicture}
    \caption{Tracking setup with scan head (1), galvo mirrors (2), lens (3), motorized reference arm (4), sample holder (5), and robot arm (6).}
    \label{f:setup}
\end{figure}

The experimental setup is illustrated in Fig.~\ref{f:setup}. We attach the samples to a 6-axis robot arm (IRB 120, ABB, Switzerland) to simulate motion. The robot arm has a repeatability of \SI{0.01}{\milli\meter} and shifts the samples along its base coordinate frame's axes. These axes are roughly comparable to the axes of the OCT data and the axes defined by the galvos and the stepper motor of the reference arm. However, a lateral motion will always lead to some small axial compensations and vice versa. 

For a quantitative evaluation of the tracking accuracy, we need a calibration between the coordinate frame of the robot arm and the coordinate frame defined by the tracking system. We move the robot very slowly with the plate phantom attached to \num{125} positions within a grid of size \SI{40x40x40}{\milli\meter} with active tracking and compensation. To account for noise, at each grid point we log the positions of robot arm, galvos and stepper motor \num{20} times, spaced out by \SI{100}{\milli\second} in between.  While the robot's base frame is clearly Cartesian, moving the galvos shifts the C-scan on a slightly spherical path. Therefore, we first determine an affine transformation which roughly aligns the coordinate frames. Afterwards, we determine the coefficients of a three-dimensional quadratic function which mainly compensates for the remaining spherical distortions. With these two transformations, we evaluate the tracking errors in the coordinate frame of the robot in physical units. 

During the experiments, we log the positions of the robot arm and the positions of our galvos and the stepper motor at a rate of \SI{83}{\hertz}. We do not apply any temporal synchronization but rely on the logged timestamps. Each tracking experiment is recorded for \SI{60}{\second} with active motion by the robot arm and is repeated six times for each maximum velocity setting with different templates. The samples are moved linearly back and forth along a distance of about \SI{30}{\milli\meter}. For lateral motion, the robot arm moves along a diagonal which requires both galvos to compensate the motion.  If the tracking error is above \SI{2}{\milli\meter} for at least one pair of logged position, we define that tracking failed.

\section{RESULTS AND DISCUSSION}

The calibration between robot arm and tracking system leads to a residual root-mean-square error (RMSE) of \SI{0.11}{\milli\meter}. This order of error is expectable in the current setup due to the spacing of the A-scans in the C-scans and our tracking algorithm which does not provide sub-pixel translations.

\begin{figure}[bt!]
    \centering
    \subfloat[Tracking of lateral motion failed once for the plate at \num{40}, once at \num{70}, and four times at \SI{85}{\milli\meter\per\second}. For turkey, it failed twice at \num{40}, twice at \num{55}, and only succeeded twice at \num{70} and once at \SI{85}{\milli\meter\per\second}. \label{f:results:lateral}]{
        \includegraphics[width=.95\linewidth,height=0.19\textheight]{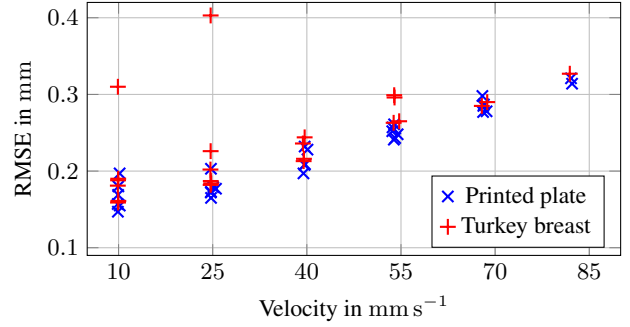}
    }\\
    \subfloat[Tracking of axial motion failed once for the plate at \num{30} and once at \SI{40}{\milli\meter\per\second}. For turkey, it failed once at \num{35} and once at \SI{40}{\milli\meter\per\second}.
    \label{f:results:axial}]{
        \includegraphics[width=.95\linewidth,height=0.19\textheight]{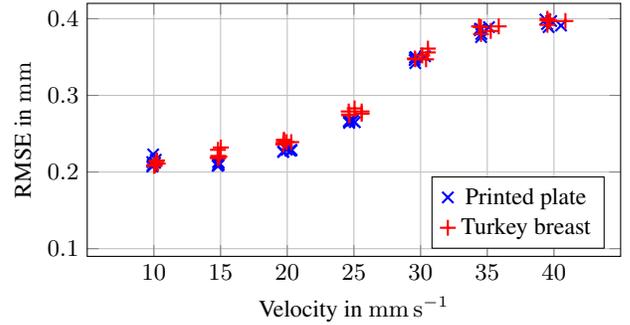}
    }\\
    \subfloat[Tracking of 3D motion failed once for the plate at \num{15}, once at \num{35}, twice at \SI{40}{\milli\meter\per\second}. For turkey, it failed once at \num{20}, once at \num{30}, thrice at \num{35}, and there was only one success at \SI{40}{\milli\meter\per\second}.
    \label{f:results:full}]{
        \includegraphics[width=.95\linewidth,height=0.19\textheight]{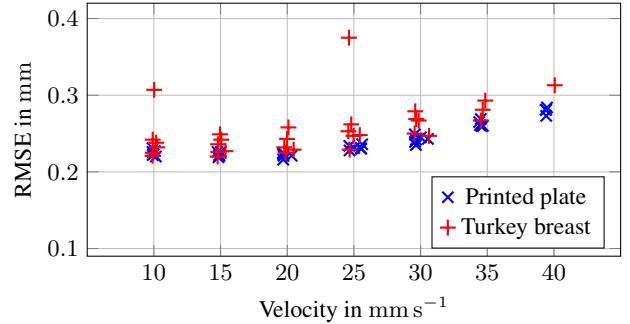}
    }
    \caption{RSME values for tracking the plate and turkey samples at different velocities. We evaluate solely lateral (a), solely axial (b), and fully three-dimensional motion (c).}
    \label{f:results}
\end{figure}

First, we focus on the results in Fig.~\ref{f:results:lateral} for lateral motion, which is mainly compensated by the two galvos. For \SI{10}{\milli\meter\per\second} the RMSE remains below \SI{0.2}{\milli\meter} except for one area of a turkey sample. This area turns out to be difficult for tracking and leads to many failures or exceptionally high errors throughout all measurements. In practice,  however, one would rather test beforehand whether a region is suitable. The RMSE increases for \SI{85}{\milli\meter\per\second} to slightly above \SI{0.3}{\milli\meter}.
 
\begin{figure}[tb!]
    \centering
    \includegraphics[width=.95\linewidth,height=0.18\textheight]{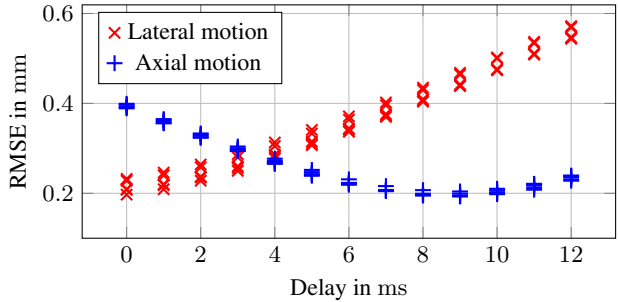}
    \caption{RSME values for the plate moving axially or laterally with \SI{40}{\milli\meter\per\second} for different delays between the robot arm and our tracking system.}
    \label{f:latency}
\end{figure}

For axial motion, which is mainly compensated by the motorized reference arm, Fig.~\ref{f:results:axial} shows that we can track lower velocities. If we temporally shift the logged positions, as shown in Fig.~\ref{f:latency}, we observe that the increased error compared to lateral tracking is caused by a delay of the stepper motor. This delay increases with the velocity of the tracked sample.

Finally, Fig.~\ref{f:results:full} shows the results for simultaneous linear motion along all three axes. The results are comparable to those for only axial motion. However, the galvos can partially compensate the worse performance of the stepper motor, because the effective motion on each individual is now smaller than for pure axial motion. Thereby, the RMSE stays below \SI{0.3}{\milli\meter} for the plate and is only slightly higher for the turkey samples. Again, one template volume of a turkey sample produces extraordinary high errors and tracking even failed for rather slow velocities for this template.

\section{Conclusion}
We presented a hardware and software setup based on OCT that is able to follow motion along all spatial directions reliably for up to \SI{25}{\milli\meter\per\second} with RMSEs in the order of \SI{0.25}{\milli\meter}.  The motion distance of \SI{30}{\milli\meter} was substantially larger than the OCT's FOV but the template was lost only once for these velocities during experiments which consisted of \SI{60}{\second} of motion. Even \SI{85}{\milli\meter\per\second} could be successfully tracked laterally which shows the potential and is more than four times faster than for the previously proposed setup. Furthermore, our processing pipeline increased the data rate by more than one order of magnitude by realizing a processing rate of \SI{831}{\hertz} which is also promising, even for applications like haptic feedback. Alternatively, one could reduce the tracking rate and instead use volumes with higher lateral resolution. Nevertheless, the motorized reference arm is currently the limiting factor. Stepper motors are not designed for this kind of use and a different approach should be considered if higher velocities and frequent changes of direction occur. Furthermore, there is a trade-off between the accuracy and the maximum trackable velocity. In this work, we focused on the latter aspect and tuned the system accordingly. 

\bibliographystyle{IEEEbib}
\bibliography{refs}

\pagebreak

\end{document}